\documentclass[12pt]{article}
\usepackage{times}
\usepackage{geometry}
\usepackage[utf8]{inputenc}
\usepackage{enumitem,amssymb}
\usepackage{ragged2e}
\newlist{thematic}{itemize}{8}
\setlist[thematic]{label=$\square$}
\usepackage{pifont}
\usepackage{natbib}
\usepackage{psfig}
\usepackage{wrapfig}
\usepackage{graphicx}

\begin{document}
\begin{flushleft}
\huge
Astro2020 APC White Paper \linebreak

The Compton Spectrometer and Imager \linebreak
\normalsize

\thispagestyle{empty}

\textbf{Principal Author:}

Name:	John A. Tomsick
 \linebreak						
Institution:  UC Berkeley
 \linebreak
Email: jtomsick@berkeley.edu
 \linebreak
 
\textbf{Co-authors:} Andreas Zoglauer (UCB), Clio Sleator (UCB), Hadar Lazar (UCB), Jacqueline Beechert (UCB),  Steven Boggs (UCSD and UCB), Jarred Roberts (UCSD), Thomas Siegert (UCSD), Alex Lowell (UCSD), Eric Wulf (NRL), Eric Grove (NRL), Bernard Phlips (NRL), Terri Brandt (GSFC), Alan Smale (GSFC), Carolyn Kierans (GSFC), Eric Burns (GSFC), Dieter Hartmann (Clemson), Mark Leising (Clemson), Marco Ajello (Clemson), Chris Fryer (LANL), Mark Amman (independent), Hsiang-Kuang Chang (NTHU, Taiwan), Pierre Jean (IRAP, France), \& Peter von Ballmoos (IRAP, France)

\end{flushleft}

\begin{figure}[h]
\begin{center}
\includegraphics[height=2.8in,angle=0]{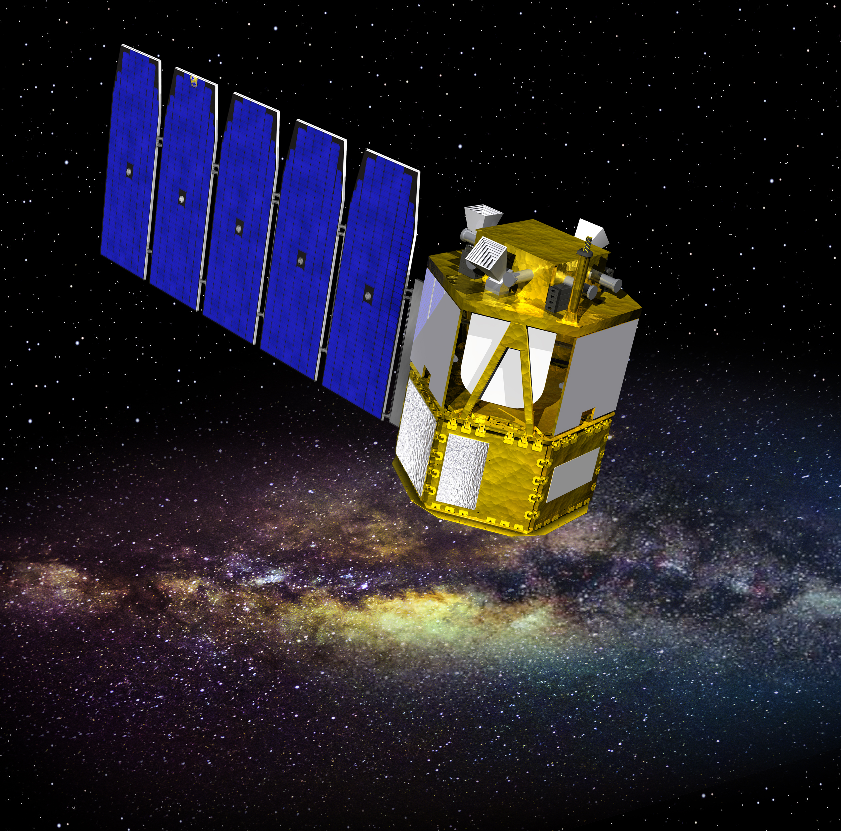}
\end{center}
\end{figure}

\noindent
\textbf{Summary:}
In this Astro2020 APC White Paper, we describe a Small Explorer (SMEX) mission concept called the {\em Compton Spectrometer and Imager}.  {\em COSI} is a Compton telescope that covers the bandpass often referred to as the ``MeV Gap" because it is the least explored region of the whole electromagnetic spectrum.  {\em COSI} provides a significant improvement in sensitivity along with high-resolution spectroscopy, enabling studies of 511\,keV electron-positron annihilation emission and measurements of several radioactive elements that trace the Galactic history of supernovae.  {\em COSI} also measures polarization of gamma-ray bursts (GRBs), accreting black holes, and pulsars as well as detecting and localizing multimessenger sources.  In the following, we describe the {\em COSI} science, the instrument, and its capabilities.  We highlight many Astro2020 science WPs that describe the {\em COSI} science in depth.


\pagebreak

\section{Overview}

\setcounter{page}{1}

The {\em Compton Spectrometer and Imager (COSI)}\footnote{Also see http://cosi.ssl.berkeley.edu} is a wide-field-of-view telescope designed to survey the gamma-ray sky at 0.2--5\,MeV, performing high-resolution spectroscopy, wide-field imaging, and polarization measurements (see Table~\ref{tab:performance} for performance parameters). {\em COSI} utilizes the most effective spectroscopy and proven polarization technology in this energy range. This APC white paper describes a {\em COSI} design for a Small Explorer mission (called {\em COSI-SMEX} in some Astro2020 science white papers) and the capabilities of such a mission.

{\em COSI} will map the Galactic positron annihilation emission, revealing the mysterious concentration of this emission near the Galactic center and the Galactic bulge in unprecedented detail.  {\em COSI} will elucidate the role of supernovae (SNe) and other stellar populations in the creation and evolution of the elements by mapping tracer elements. {\em COSI} will image $^{26}$Al with unprecedented sensitivity, perform the first mapping of $^{60}$Fe, search for young, hidden supernova remnants through $^{44}$Ti emission, and enable a host of other nuclear astrophysics studies. {\em COSI} will also study compact objects both in our Galaxy and AGN as well as gamma-ray bursts (GRBs), providing novel measurements of polarization as well as detailed spectra and light curves. The {\em COSI} field of view and localization capabilities make it powerful for searches of electromagnetic counterparts to gravitational wave and high-energy neutrino detections.  {\em COSI} will also address science topics related to the cosmic MeV background as well as the Galactic gamma-ray continuum, establishing the energy distribution of high-energy cosmic rays.

\begin{wrapfigure}{r}{4.0in}
\vspace{-0.3cm}
\centerline{\includegraphics[height=2.6in,angle=0]{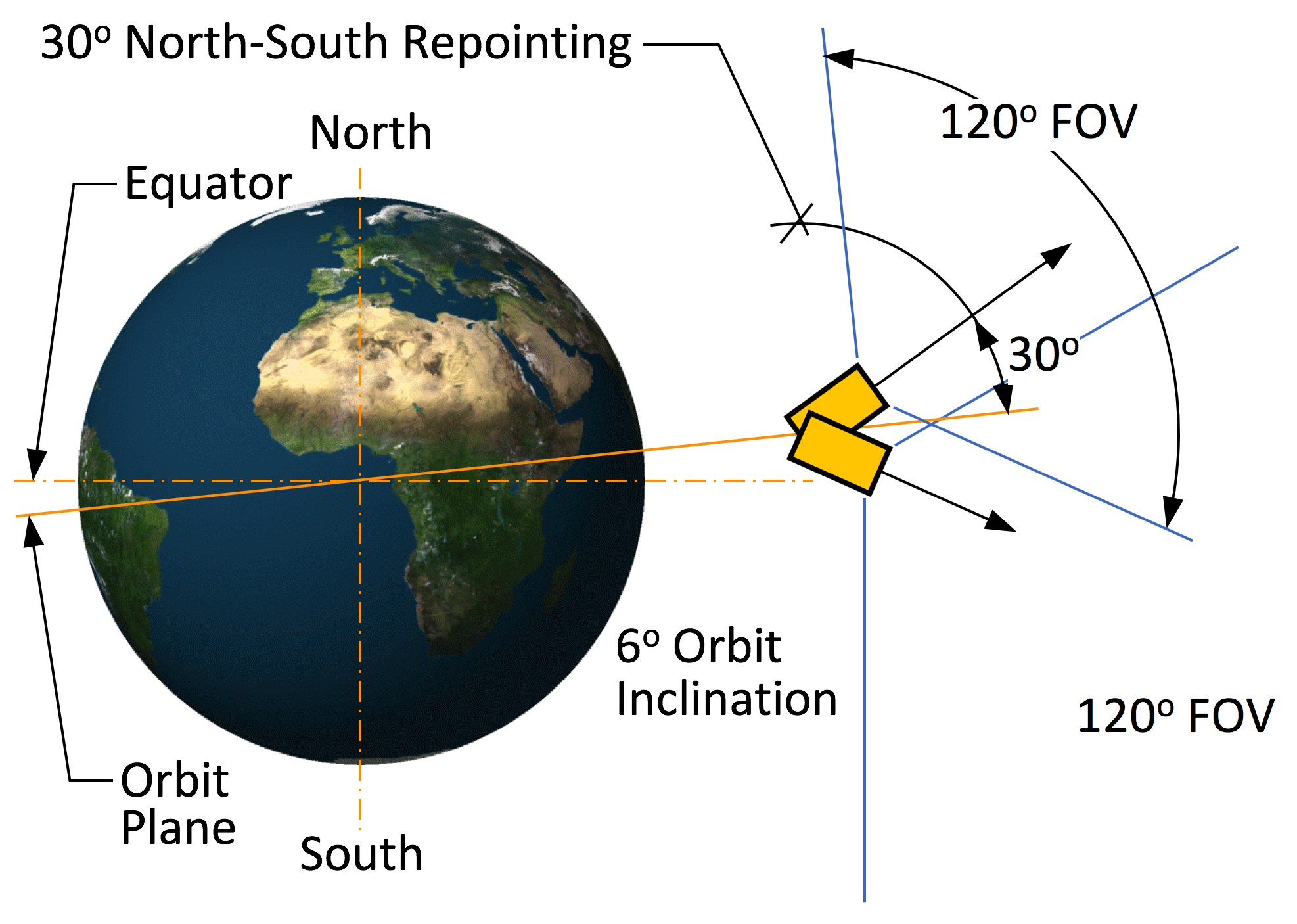}}
\vspace{-0.5cm}
\caption{\small {\em COSI} (yellow rectangles) is shown in a low-inclination low-Earth orbit.  In survey mode, it monitors the entire sky on daily time scales via North-South repointing.  During the prime mission, {\em COSI} would spend more than 90\% of its time in survey mode. It also has an inertial pointing mode for targets of opportunity.\label{fig:mission_overview}}
\end{wrapfigure}

The heart of {\em COSI} is a stacked array of germanium cross-strip detectors, which provide high efficiency, high resolution spectroscopy, and precise 3D positioning of photon interactions. As {\em COSI} is a Compton telescope, 3D positioning is required to carry out the Compton event reconstruction. We accomplish this using the versatile MEGAlib software package \citep{zoglauer06}.  The detectors are housed in a cryostat and cryogenically cooled to $<$80\,K.  They are shielded on five sides, reducing the background and defining the field of view to be 25\% of the sky.  The {\em COSI} shields are active scintillators, extending {\em COSI}'s field of view for detection of GRBs and other sources of gamma-ray flares to approximately 50\% of the sky.  {\em COSI} will be in low-Earth orbit and will spend most of the mission time in a survey mode, alternately pointing $30^{\circ}$ North of zenith to $30^{\circ}$ South of zenith every 12 hours to enable complete sky coverage over a day (see Figure~\ref{fig:mission_overview}).


\begin{table}[t]
\caption{Basic Performance Parameters\label{tab:performance}}
\begin{minipage}{\linewidth}
\begin{center}
\small
\begin{tabular}{cc|cc}
\hline \hline
{\bf Parameter} & {\bf {\em COSI} value} & {\bf Parameter} & {\bf {\em COSI} value}\\ \hline
Energy Range & 0.2-5\,MeV & Spectral Resolution & 0.2-1\%\\ \hline
Field of View & 25\% sky & Sky Coverage & 100\% per day\\ \hline
Angular Resolution (511\,keV) & 3.2$^{\circ}$ FWHM & Angular Resolution (1.809\,MeV) & 1.5$^{\circ}$ FWHM\\ \hline
Line Sensitivity\footnote{3$\sigma$ narrow line sensitivity in 2-years of survey observations in units of photons\,cm$^{-2}$\,s$^{-1}$.}  (511\,keV) & $7.9\times 10^{-6}$ & Line Sensitivity$^{a}$ (1.809 MeV) & $1.7\times 10^{-6}$\\ \hline
Flux limit for polarization & 15\,mCrab & Fluence limit for GRB polarization\footnote{50\% minimum detectable polarization in units of erg\,cm$^{-2}$.} & $4\times 10^{-6}$\\ \hline
\end{tabular}
\end{center}
\end{minipage}
\end{table}

\section{Key Science Goals and Objectives}

\begin{figure}[h]
\begin{center}
\includegraphics[height=2.5in,angle=0]{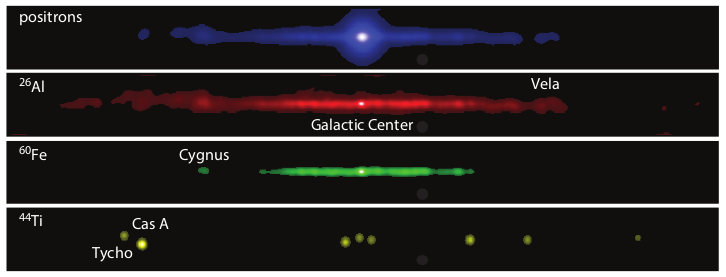}
\end{center}
\vspace{-0.5cm}
\caption{\small
The Radioactive Milky Way.  The images are {\em COSI} simulations for the entire Galactic plane ($l = \pm 180^{\circ}$ and $b = \pm 15^{\circ}$).  The simulated positron map is based on the bulge measured by {\em INTEGRAL}/SPI and the 240$\mu$m map as a tracer for the disk.  The $^{26}$Al (1.809 MeV) and $^{60}$Fe (1.173/1.333 MeV) maps also use the 240$\mu$m map and fluxes consistent with measurements by {\em COMPTEL} and SPI.  In contrast to the $\sim$Myr half lives of $^{26}$Al and $^{60}$Fe, the short, 60 yr $^{44}$Ti (1.157 MeV) half-life traces recent supernova events.}\label{fig:radioactive_mw}
\end{figure}

{\em COSI} has a large science portfolio that includes all-sky gamma-ray imaging for mapping electron-positron annihilation emission, radioactive decay lines, and Galactic diffuse emission. The long radioactive decay time scales of $^{26}$Al and $^{60}$Fe provide a look at where in the Galaxy nucleosynthesis has occurred over the past millions of years and where it is occurring now. While $^{60}$Fe is almost exclusively released into the interstellar medium (ISM) during core collapse supernovae (CCSNe), $^{26}$Al is also thought to be produced (and released) in the winds of massive stars.  $^{44}$Ti produced in SNe decays quickly by comparison, showing where recent supernova events have occurred in the last several hundred years and allowing for detailed studies of individual supernova remnants. The decay of $^{26}$Al, which produces a 1.809\,MeV gamma-ray and a positron, is the prominent known source of positrons through nucleosynthesis, but observations of the 0.511\,MeV positron annihilation line indicate that one or more additional unknown sources of positrons dominate their production. As illustrated in Figure~\ref{fig:radioactive_mw} and discussed in detail below, it is by comparing the full Galaxy maps of positron annihilation and radioactive isotopes from nucleosynthesis that we can understand the origin of the ``extra" positrons and obtain a more complete picture of nucleosynthesis in the Galaxy. The high sensitivity that {\em COSI} provides for these emission lines compared to earlier measurements (Figure~\ref{fig:lin_sens}) makes this goal possible.

\subsection{Probing the Origin of Galactic Positrons}

The production of positrons and their annihilation in the Galactic ISM is one of the pioneering topics of gamma-ray astronomy. Despite five decades of study since the initial detection of the 511\,keV line from the inner Galaxy \citep{jhh72,lms78}, the origin of these positrons remains uncertain. {\em INTEGRAL}/SPI's imaging led to a map of the annihilation emission from the Galaxy that shows a bright bulge and fainter disk \citep{bouchet10}.  The angular resolution of the SPI map does not provide any definitive information about sub-structure in the emission.  No individual sources have been detected, and the origin of the bright bulge emission is unclear. The scale height of the disk emission is also uncertain, with \cite{skinner14} finding an rms height of $\sim$$3^{\circ}$, while \cite{siegert16} obtain $>$$9^{\circ}$.  This raises questions about the origin of the positrons as well as how far they propagate.  Even more fundamentally, the uncertain scale height means that the number of positrons in the disk is uncertain by a factor of more than three.  Thus, even though the 511\,keV line is the strongest persistent gamma-ray signal, fundamental gaps remain in our understanding of this anti-matter component of our Milky Way.

The Astro2020 WP entitled, ``Positron Annihilation in the Galaxy," by \cite{kierans19} focuses on the positron science that can be addressed with a sensitive wide FoV imager in the MeV band with excellent energy resolution. The specific science goals discussed include: determining whether the 511\,keV emission is truly diffuse or whether there are individual sources; constraining the positron propagation distance by comparing the $^{26}$Al (1.809\,MeV) distribution as well as other source distributions (e.g., pulsars) to the 511\,keV distribution; probing the conditions in different regions of the Galaxy where positron annihilation occurs; and measuring or placing limits on the injection energy of positrons into the ISM from measurement of the MeV continuum due to annihilation in flight. This will constrain the mass of a possible contributing dark matter particle, as well as the contributions of black holes and pulsars.

{\em COSI}'s capabilities (see Table~\ref{tab:performance}) are well-matched to these goals.  The excellent spectral resolution provides a leap in sensitivity and also allows for measurements of emission line shapes (e.g., width of the 511\,keV line components, Doppler shifts of $^{44}$Ti).  The angular resolution will allow for a sensitive search for point sources and will also easily distinguish between a disk scale height of $3^{\circ}$ and $>$9$^{\circ}$.  In addition to constraints on positron propagation, {\em COSI}'s measurements at 511\,keV and 1.809\,MeV will allow us to determine what fraction of the positrons are accounted for by $^{26}$Al decay.

\subsection{Revealing Element Formation}

The MeV bandpass includes several nuclear emission lines that probe different physical processes in our Galaxy and beyond.  Long-lived isotopes such as $^{26}$Al (1.809\,MeV line) and $^{60}$Fe (1.173 and 1.333\,MeV lines), predominantly produced in SNe, provide information about the galaxy-wide star formation history, integrated over the past million years.  To first order, images of the Galaxy at these energies trace the last $\sim$10,000 CCSNe.  $^{44}$Ti (1.157\,MeV as well as 68 and 78\,keV lines in the hard X-ray range), with a half-life of 60 years, traces young Galactic SNe which occurred in the last few hundred years, and $^{56}$Co (0.847 and 1.238\,MeV) decays so rapidly (half-life: 77 days) that it is currently only seen by following up SNe in nearby galaxies. These lines allow us to trace the amount and distribution of these radioactive isotopes created in the underlying SNe giving us an independent and more direct view of Galactic chemical evolution. 

\begin{figure}[t]
\begin{center}
\includegraphics[height=3.5in,angle=0]{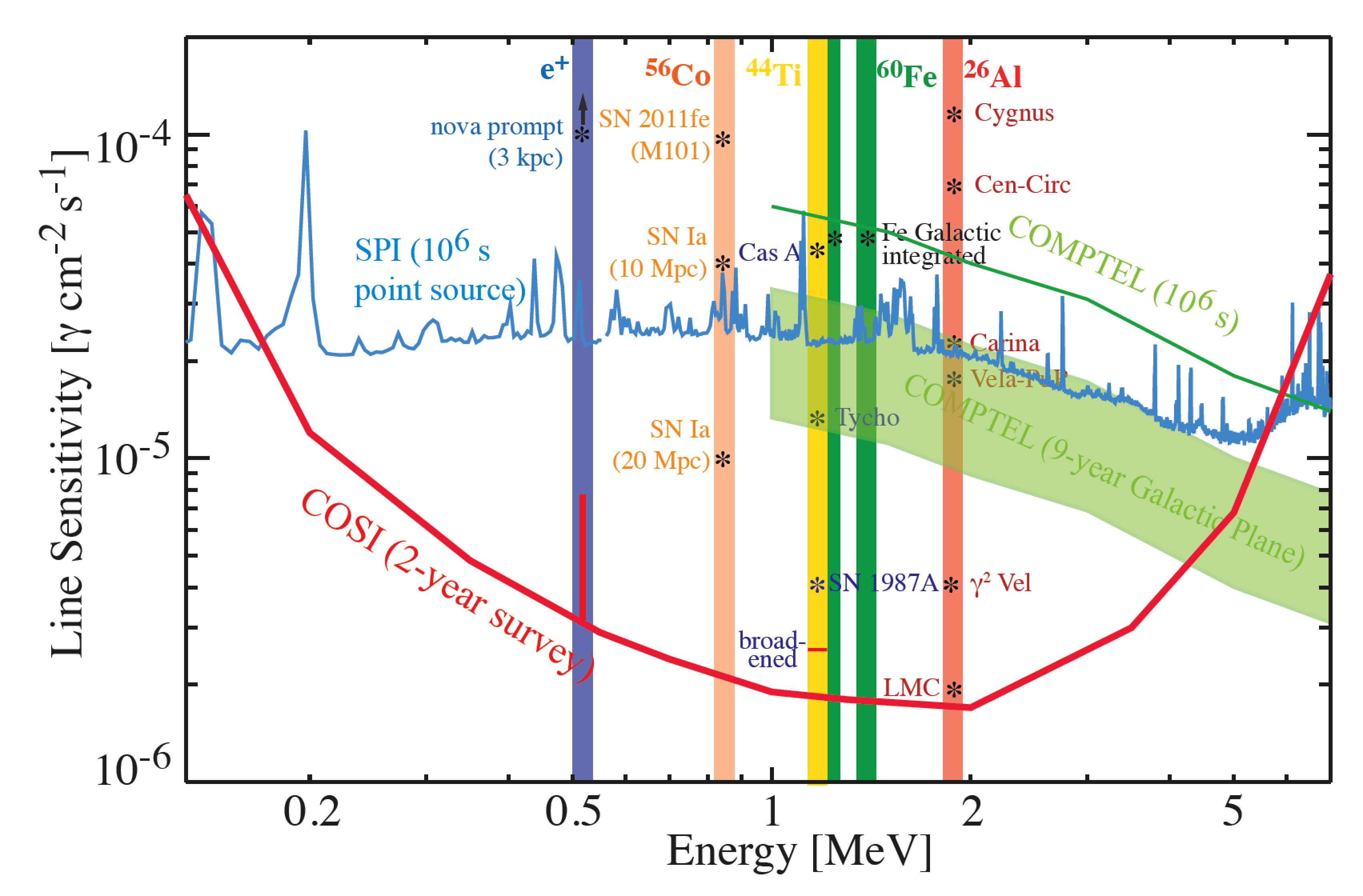}
\end{center}
\vspace{-0.5cm}
\caption{\small The {\em COSI} narrow line sensitivity (3$\sigma$) compared with {\em COMPTEL} and {\em INTEGRAL}/SPI.}\label{fig:lin_sens}
\end{figure}

The Astro2020 WP entitled, ``Catching Element Formation in the Act. The Case for a New MeV Gamma-Ray Mission: Radionuclide Astronomy in the 2020s," by \cite{timmes19} describes the scientific opportunities in detail.  They include learning about the white dwarf progenitors of Type Ia SNe by studying the time evolution of the $^{56}$Co line, understanding how massive stars explode by studying the properties of the $^{44}$Ti emission from remnants of core-collapse SNe, and understanding how newly synthesized elements from massive stars, SNe, and novae enrich the Galaxy using $^{26}$Al and $^{60}$Fe as tracers.

{\em COSI} provides unprecedented sensitivity to these emission lines as shown in Figure~\ref{fig:lin_sens}.  It will carry out a Galaxy-wide search for $^{44}$Ti sources, provide the first map of $^{60}$Fe (to date, the only detection has been obtained for the full Galaxy combined), and greatly improve on the $^{26}$Al maps measured by COMPTEL and SPI.  As $^{60}$Fe is released into the ISM only when CCSNe occur while $^{26}$Al is also released through the winds of massive stars, the ratio of emission from these two isotopes provides detailed and unique information about massive star evolution, the supernova rate, the star formation history, the ages of OB associations, and the formation of black holes and neutron stars in different regions of the Galaxy.

\subsection{Insight into Extreme Environments with Polarization}

Polarization measurements provide unique diagnostics for determining emission mechanisms and source geometries (e.g., magnetic field, accretion disk, and jet), but current results give just a glimpse into the potential for what we can learn.  In X-rays, the Crab nebula was detected at a level of 19\% \citep{novick72,weisskopf78}, but most sources show upper limits well below this, requiring a very sensitive X-ray polarimeter like the upcoming Imaging X-ray Polarimetry Explorer \citep[IXPE][]{weisskopf16}.  Above 200\,keV, in the {\em COSI} bandpass, much higher polarization levels have been measured for the Crab \citep{dean08,forot08} and for the accreting black hole Cygnus~X-1 \citep{laurent11,jourdain12}. 

A Compton telescope like {\em COSI} has intrinsic polarization capability because the gamma-rays preferentially scatter in a direction perpendicular to their electric field.  Thus, with the sensitivity of {\em COSI} to polarization and the fact that {\em COSI} observes in a bandpass where the polarization level can be expected to be high in several cases, {\em COSI} has a large discovery space.  In addition to pulsars (like the Crab) and accreting stellar mass black holes (like Cyg~X-1), {\em COSI} will be able to measure polarization for several AGN (like Cen~A), and its wide field of view allows for detection of large numbers of GRBs, for which polarization levels above 50\% have been reported.  In a 2-year mission, we expect {\em COSI} to see at least 40 GRBs with fluences in excess of $4\times 10^{-6}$\,erg\,cm$^{-2}$, which corresponds to the limit for obtaining a polarization measurement with a minimum detectable polarization of 50\%.

The Astro2020 WP entitled, ``Prompt Emission Polarimeter of Gamma-Ray Bursts," by \cite{mcconnell19} discusses the promise of gamma-ray polarimetry for constraining the GRB jet magnetic field structure, the jet composition, the energy dissipation process, and the emission mechanism.  For AGN, the WP entitled, ``High-Energy Polarimetry - a new window to probe extreme physics in AGN jets," by \cite{rani19} describes the potential for using gamma-ray polarimetry to probe AGN physics.

\subsection{Multimessenger Astrophysics}

{\em COSI} significantly contributes to multimessenger astrophysics (MMA) with its capability to find counterparts to multimessenger sources. This includes counterparts to gravitational wave events from binary neutron star (BNS) mergers and possibly even stellar mass binary black hole mergers \citep{shawhan19} as well as neutrino detections. 

For gravitational waves from binary neutron star (BNS) mergers like GW170817/GRB170817A \citep{abbott17b,goldstein17}, {\em COSI} is able to detect and localize the BNS by finding the associated short GRB.  Scintillator-based instruments have large fields of view but poor localization capabilities, and coded aperture mask instruments have good localization capabilities but smaller fields of view than {\em COSI}.  With Compton imaging, {\em COSI} combines the capability for sub-degree localizations (or much better for bright GRBs) with instantaneous coverage of 25\% of the sky to fulfill an important need for gamma-ray capabilities to detect and localize GRBs.  We estimate detection of $\sim$15-20 short GRBs in two years.  While the COSI shields will only provide rough localization, they double the {\em COSI} field of view and allow comparison of the arrival time of short GRB to the arrival of the gravitational wave signal from BNS merger. While the cause of the 1.7\,sec delay of GRB170817A is uncertain, it may relate to the time for internal shocks to form in the jet, the shock breakout time, or the time for the merging NSs to collapse into a BH.  The Astro2020 WPs entitled, ``A Summary of Multimessenger Science with Neutron Star Mergers," by \cite{burns19a} and ``Gamma-Rays and Gravitational Waves," by \cite{burns19b} describe the potential groundbreaking science associated with BNSs and the important role that gamma-ray observations play.

The other recent development in MMA was the likely association of a high-energy neutrino with the blazar TXS 0506+056 \citep{icecube18}. The primary evidence for the association was the simultaneous gamma-ray flaring from the blazar at the time of the high-energy neutrino detection.  {\em COSI} has the sensitivity to detect such gamma-ray flares and localize the flaring source for future high-energy neutrino events.  
The Astro2020 WPs entitled, ``A Unique Messenger to Probe Active Galactic Nuclei: High-Energy Neutrinos," by \cite{santander19} describes prospects for further advances in this area.

\subsection{Extended Science Portfolio}

{\em COSI} will also contribute to the science described in several other Astro2020 WPs:\\
\noindent
$\bullet$~``Thermonuclear Supernovae" by \cite{zingale19} is on multimessenger observations of Type Ia SNe, which are caused by the thermonuclear explosion of approximately a solar mass of material triggered by accretion onto a white dwarf or merging white dwarfs.  The WP includes a discussion of what can be learned from spectroscopy in the MeV band.\\
\noindent
$\bullet$~Classical Novae are also caused by thermonuclear explosions on the surface of white dwarfs but with less mass than for Type Ia SNe.  A wide FOV gamma-ray spectrometer would provide the first opportunity to measure short-lived emission lines such as the 478\,keV line from $^{7}$Be, the 1.275\,MeV line from $^{22}$Na, and the 511\,keV flash from $^{13}$N and $^{18}$F.  Novae are also known to produce gamma-ray continuum emission as described in the WP by \cite{chomiuk19}.\\
\noindent
$\bullet$~``The MeV Background" \citep{ajello19a} and ``Supermassive Black Holes at High Redshifts" \citep{ajello19b}  both discuss improving MeV sensitivity to detect fainter AGN or blazars. Such improvement would allow better determination of the nature of the MeV background and more in-depth studies of the early Universe.\\
\noindent
$\bullet$~``Energetic Particles of Cosmic Accelerators" includes science of particle acceleration and non-thermal processes enabled by gamma-ray observations of Galactic sources \citep{venters19a} and extragalactic sources \citep{venters19b}.\\
\noindent
$\bullet$~Although rare, it is imperative to have a gamma-ray spectrometer in orbit when the next supernova occurs in the Galaxy or in a nearby galaxy (e.g., LMC, SMC) as described in the ``Core-Collapse Supernovae and Multi-Messenger Astronomy" WP by \cite{fryer19}.  Gamma-ray emission lines probe both the nature of shell burning in the stellar prgogenitor and the SN explosion properties.\\

\section{Technical Overview}

\begin{wrapfigure}{r}{3.0in}
\vspace{0cm}
\centerline{\includegraphics[height=2.2in,angle=0]{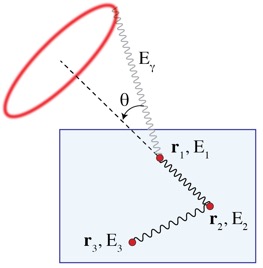}}
\vspace{0.0cm}
\caption{\small Schematic of interactions for a single gamma-ray in a Compton telescope.
\label{fig:schematic}}
\end{wrapfigure}

The {\em COSI} instrument utilizes Compton imaging of gamma-ray photons \citep{vonballmoos89,bj00}.  An incoming photon of energy $E_{\gamma}$ undergoes a Compton scatter at a polar angle $\theta$ with respect to its initial direction at the position $r_{1}$, creating a recoil electron of energy $E_{1}$ that induces the signal measured in the detector (see Figure~\ref{fig:schematic}). The scattered photon then undergoes a series of one or more interactions, which are also measured. The Compton formula relates the initial photon direction to the scatter direction (measured direction from $r_{1}$ to $r_{2}$) and the energies of the incident and scattered gamma-rays. 

\begin{figure}[t]
\begin{center}
\includegraphics[height=3.8in,angle=0]{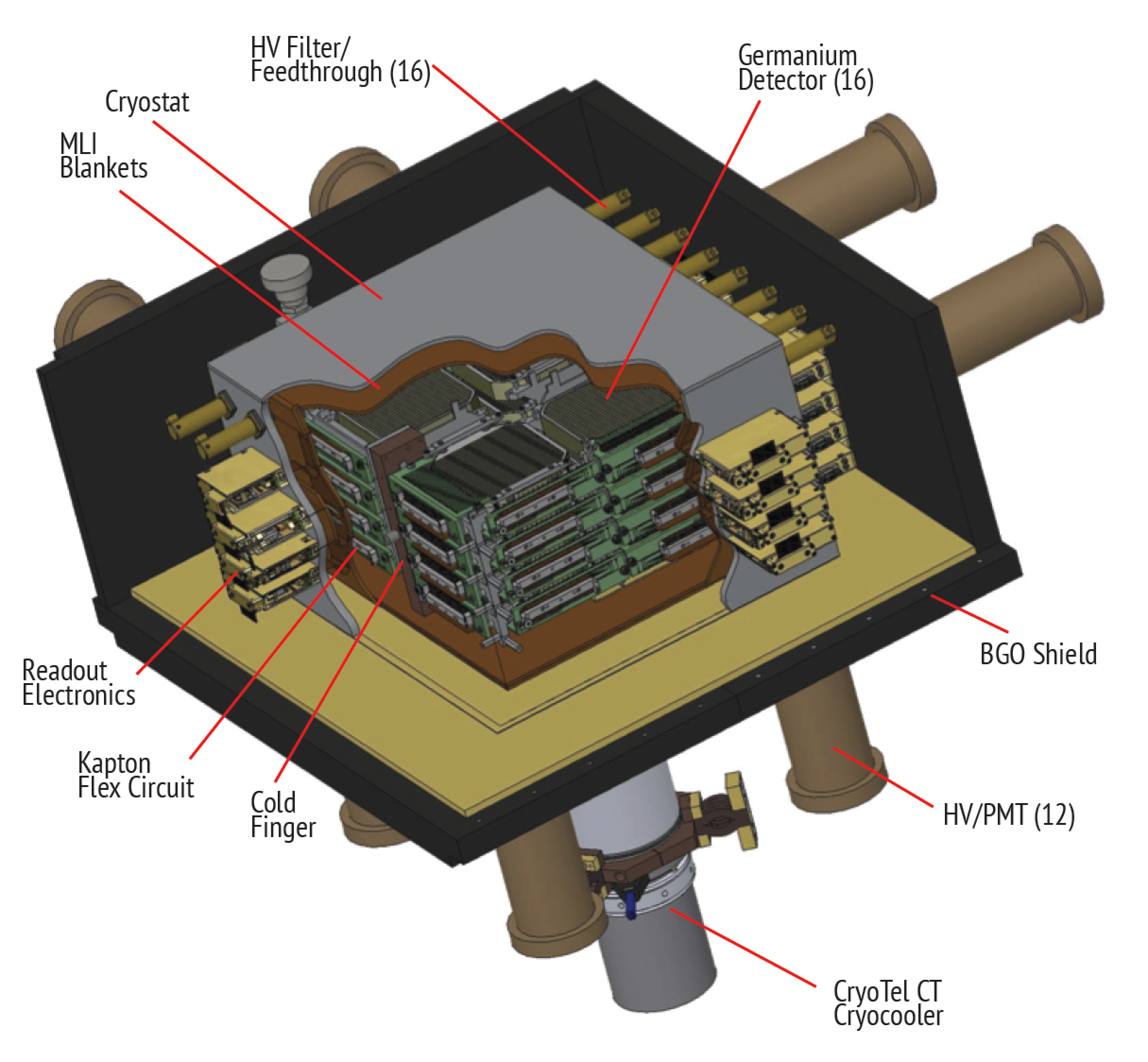}
\end{center}
\vspace{-0.5cm}
\caption{\small 
Cutaway view of the {\em COSI} instrument.  Each germanium detector is $8\times8\times1.5$\,cm$^{3}$.  The BGO shield box, shown without mounting brackets and materials, is $41\times41\times14$\,cm$^{3}$.
\label{fig:cosi_instrument}}
\end{figure}

{\em COSI} employs a novel Compton telescope design using a compact array of cross-strip germanium detectors (GeDs) to resolve individual gamma-ray interactions with high spectral and spatial resolution. No other technology thus far tested exceeds {\em COSI}'s spectral resolution and polarization capability in this energy range. The {\em COSI} array of 16 GeDs is housed in a common vacuum cryostat cooled by a CryoTel CT mechanical cryocooler. The GeDs are read out by a custom ASIC, which is integrated into the full data acquisition system. An active BGO shield encloses the cryostat on the sides and bottom to veto events from outside the FOV.  The {\em COSI} instrument is shown in Figure~\ref{fig:cosi_instrument}.

We have developed {\em COSI} through NASA's Astrophysics Research and Analysis (APRA) program, including high-altitude balloon flights in 2005 and 2009 under the {\em Nuclear Compton Telescope (NCT)} name and in 2014 and 2016 as {\em COSI}.  Milestones include the detection of the Crab Nebula in 2009 \citep{bandstra11} and a very successful superpressure balloon flight in 2016.  This flight took off from New Zealand and lasted for 46 days before landing in Peru.  The data we obtained with {\em COSI-APRA} provides a strong proof of concept for {\em COSI}.  During the flight, we were able to detect, localize, and report GRB 160530A in real time \citep{tomsick16_cosi}.  An early summary of the results from the full flight (detection of the Crab, Cyg~X-1, Cen~A, the GRB, and the 511\,keV emission from the Galactic Center) is reported in \cite{kierans17}.  An analysis of the GRB 160530A data resulted in an upper limit on the polarization of $<$46\% \citep{lowell17a}, and new techniques were developed for polarization studies \citep{lowell17b} and also for calibrating the instrument \citep{lowell17_phd,yang18}.  The detection of the 511\,keV emission and information about the line shape and spatial extent is reported in \cite{kierans18_phd}. We have also made progress in characterizing the detailed detector response (Sleator et al., submitted to NIM A) and the response of the overall instrument for spectral analysis (Sleator et al., PhD thesis, submitted to UC Berkeley thesis committee).


\section{Schedule, Cost, and Current Status}

We are currently in the final phases of preparing a SMEX proposal for the 2019 Astrophysics call.  Our instrument concept is mature, and the schedule including margin allows launch in 2025. {\em COSI} fits within the SMEX envelope for cost (\$145M including margin), mass, and power. For a mission focusing on the 0.2-5\,MeV energy range with high spectral resolution and wide field of view, {\em COSI} as a SMEX is a natural next step. Thus, continued support for projects in the Astro2020 category of Small ($<$\$500M) Space-Based Projects will enable this exciting science.

\section{Organization and Partnerships}

The project is a collaboration between UC Berkeley’s Space Sciences Laboratory, UC San Diego, the Naval Research Laboratory, and Goddard Space Flight Center, and includes science partners at Clemson University, Los Alamos National Laboratory, National Tsing Hua University, and the Institut de Recherche en Astrophysique et Planétologie.  We are partnering with Northrop Grumman Innovation Systems (NGIS) for the spacecraft.


\end{document}